\documentclass{article}

\usepackage{times}
\usepackage{xcolor}
\usepackage{soul}
\usepackage[utf8]{inputenc}
\usepackage[small]{caption}
\usepackage{floatrow}
\usepackage{subcaption}
\usepackage{hyperref}
\usepackage{stmaryrd}
\usepackage{amsmath}%
\usepackage{amssymb}%
\usepackage{amsthm}
\usepackage{amsfonts}
\usepackage{graphics}
\usepackage{gastex}
\usepackage{multirow}
\usepackage{cases}
\usepackage{algpseudocode}
\usepackage{algorithm}
\usepackage{verbatim}
\usepackage{enumitem}
\usepackage{mathrsfs}

\newfloatcommand{capbtabbox}{table}[][\FBwidth]

\theoremstyle{plain}

\newcommand{\BibleQA}{\textsf{BibleQA}}
\newcommand{\GloVe}{\textsf{GloVe}}
\newcommand{\LSTM}{\mathsf{LSTM}}
\newcommand{\SQuAD}{\textsf{SQuAD}}
\newcommand{\wordtvec}{\textsf{word2vec}}

\title{Finding Answers from the Word of God: \\
Domain Adaptation for Neural Networks in Biblical Question Answering}

\author{
Helen Jiahe Zhao,
Jiamou Liu
\\
Department of Computer Science \\
The University of Auckland\\ New Zealand\\
jiamou.liu@auckland.ac.nz
}

\begin{document}


\maketitle

\begin{abstract}
Question answering (QA) has significantly benefitted from deep learning techniques in recent years.
However, domain-specific QA remains a challenge due to the significant amount of data
required to train a neural network. This paper studies the answer sentence
selection task in the Bible domain and answer questions by selecting relevant verses from the
Bible. For this purpose, we create a new dataset \BibleQA\ based on bible trivia questions and propose three neural network models for our task. We pre-train our models on a large-scale QA dataset, \SQuAD, and investigate the effect of transferring weights on model accuracy. Furthermore, we also measure the model accuracies with different answer context lengths and different Bible translations. We affirm that transfer learning has a noticeable improvement in the model accuracy. We achieve relatively good results with shorter context lengths, whereas longer context lengths decreased model accuracy. We also find that using a more modern Bible translation in the dataset has a positive effect on the task.

\end{abstract}

\section{Introduction}
The desire for a computer system that could answer natural language questions has been an ability that signifies artificial intelligence. 
In recent years, neural networks and machine learning have become popular approaches for question answering (QA) tasks within the natural language processing (NLP) community. One problem with this approach is the expense of creating a suitable dataset for specific domains. Machine learning works well under the assumption that training and test are from the same distribution. Therefore, the tasks that machine learning can solve is highly dependent on the dataset. As a result, most of neural
QA research predominantly uses existing datasets, and not as much work has been
done for domain-specific QA.

In this paper, we focus on the task of answer sentence selection, specifically to answer questions by selecting verses from the Bible as answers. This task takes as input a question and a context paragraph and asks for a sentence from the context that contains the answer to the given question. In our case, the context would consist of passages from the Bible.

The Bible is not only an influential literary work but is also the most important religious
document amongst Christians. So far, not much work has been done using this corpus within
QA or other NLP tasks. Even today, it is still widely
read and studied amongst both the religious and the secular community. The Bible is often
seen as a source of wisdom where people turn to seek answers to the big questions in life. A
QA system that can answer a question using passages from the Bible has the potential to be very
beneficial for the users.

A biblical QA system could be useful for non-Christians, seeking to learn more about the Bible. They might ask questions such as: ``\textsl{Who is Jesus?}'', ``\textsl{What will happen when I die?}''. The system could then output a series
of relevant verses as answers. The same system could also be useful for scholars seeking to use the Bible as
a historical or archaeological document. They could be interested in questions such as, ``\textsl{When did Babylon destroy the Jerusalem temple?}'', ``\textsl{Where was the city of Jericho located?}''. Such questions can be answered from the relevant passages that described the historical aspect of
the Bible. Finally, one of the widest use cases could be from the Christian community who
uphold the Bible as the ultimate authority for their faith and life. They could be interested in a
wide variety of questions, ranging from theological to practical. For example, ``\textsl{Is salvation by
faith or by works?}'', ``\textsl{How should I treat people that have wronged me?}'', ``\textsl{How should I pray?}''. While many of these questions can also be answered through a search
engine, the quality of results from search engines can often be questionable and answering
using passages directly from the Bible is a valuable resource from a Christian perspective.

\paragraph*{\bf Contribution.} The goal of the research is to investigate neural-based methods for answering biblical question through verse selection. (1)
Since large-scale datasets are needed for efficient learning using neural network methods, our first contribution involves the creation of a new dataset \BibleQA. The dataset consists of Biblical questions and
the corresponding verse as derived from an existing set of questions that is available
on the Internet. (2) Then, for biblical sentence selection, we design three answer selection models based on different neural network architectures. Each of these models takes as input a question and an answer verse and outputs a predicted probability of the verse containing the answer to the question. (3) Thirdly, we leverage transfer learning techniques by pre-training the models on a larger QA dataset and provide insight into the effect of domain adaptation used in QA tasks. Our experiments also reveal how changing context lengths affects the performance of answer selection, and reveals new insights regarding various Bible translation.

\section{Related Work}

 From a text retrieval perspective,
 {\em question answering} embodies the task of finding the relevant piece of text containing the answer and subsequently extracting the answer \cite{voorhees1999trec}. This view led to open-domain QA, which encompasses the majority of today's QA systems. In recent years, QA began incorporating machine learning, with the IBM Watson being one of the most famous systems \cite{ferrucci2010building}. The primary approach behind Watson is extensive data, statistical and machine
learning analysis.
Several other neural network approaches have also been explored.
Iyyer et al. used neural networks to answer quiz bowl type questions, where given a description the task is to
identify the subject being discussed \cite{iyyer2014neural}.
Kumar et al. extended simple RNN models with an attention mechanism to enable transitive reasoning and made steps
towards reasoning-based QA \cite{kumar2016ask}.
Malinowski proposed a model using both CNN and LSTM to incorporate
image recognition and QA \cite{malinowski2015ask}.

{\em Answer sentence selection} is a QA task which involves selecting the sentence that is the most likely to contain the answer. Early approaches were predominantly syntactical, using the idea that the question and answer sentence should relate
to each other loosely through syntactical transformations. Wang et al. proposed a generative
model that transforms the answers to the questions \cite{wang2007jeopardy}. Wang and Manning introduced a
probabilistic model that models tree-edit operations on dependency parse trees, making use of
sophisticated linguistics features \cite{wang2010probabilistic}. Other similar models include using dynamic programming to find the optimal tree edit sequences \cite{yao2013answer}. The main drawback of these approaches is
that they require too much feature engineering and were difficult to adapt to new domains.
Only recently, researchers started applying neural network models. Yu et al. used CNN models for answer sentence selection on the TREC benchmark \cite{yu2014deep}. Feng et al. also proposed several CNN models for answer sentence
selection task. Wang and Nyberg constructed a joint-vector based on both the question and the
answer using an LSTM model \cite{wang2015long}.

\paragraph*{\bf NLP in religious text.} Recently, works start to emerge that use NLP for religious text mining.  The Bible is a good resource for various linguistics tasks, and has been used as a resource to improve and investigate computational linguistics tasks. Hu applied unsupervised learning to analyse Proverbs and Psalms. They clustered Psalms by content and saw
how the outcome matches the literary form of the Psalms \cite{hu2012unsupervised}. Their findings mostly matched the works by biblical scholars, but have also made unique contributions that were only made possible through machine learning methods. 
Tschuggnall and Specht explored grammar-based text analysis for authorship attribution in the
Bible \cite{tschuggnall2016plagiarism}. Faigenbaum et al. used novel image processing and machine
learning algorithms for authorship detection \cite{faigenbaum2016algorithmic}. While NLP has undoubtedly given the Biblical scholars a new method for biblical analysis, the Bible in and of itself is also an invaluable corpus for computational linguistics research.  Buchler et al.  used seven English translations of the Bible to investigate the techniques behind historical text re-use detection process and examine algorithms for paraphrase detection \cite{buchler2014towards}. The Bible provides a good test bed
for paraphrase detection as there exist several different translations all stemming from the same
origin. Agi{\'c} et al. \cite{agic2015if} used the Bible to learn part-of-speech (POS) taggers for low-resource
languages such as Akawaio, Aukan, or Cakchiquel for which the Bible is only partially translated. They learned POS taggers for 100 languages, and performs much better ($20-30\%$) than state-of-the-art unsupervised POS taggers induced from Bible translations.

\paragraph*{\bf Transfer learning in NLP.} Transfer learning and domain adaptation have been very successful with cross-domain machine learning, especially when we have a lot of data in one domain but a similar domain of interest does not have enough data for learning purposes. Computer vision has primarily benefitted from transfer learning \cite{oquab2014learning,zeiler2014visualizing}.
Large-scale image data such as ImageNet \cite{deng2009imagenet} is particularly challenging to obtain and process and researchers want to make use of the existing image data as much as possible, and many have employed transfer learning in various image processing tasks. Due to the success of transfer learning in computer vision, transfer learning has been used in NLP task such as sentiment analysis \cite{blitzer2007biographies}, POS tagger \cite{blitzer2006domain}, and machine translation \cite{axelrod2011domain}. In particular,
transfer learning has recently begun to be applied in QA. Glorot et al. used transfer learning
to examine how a system trained to answer questions from one knowledge base could answer questions from another knowledge base \cite{glorot2011domain}. Yang et al. used transfer learning for question generation \cite{yang2017semi}. For applications of domain adaptation in
neural models, the most common and the most straightforward approach is to pre-train the model on the
source data and then fine-tune the parameters on the data from the target domain \cite{min2017question,wiese2017neural}. Overall, transfer learning is essential for machine learning researchers to make use of smaller datasets, and we can expect that it will be used increasingly more in the future.

\section{\BibleQA: Bible Question-Answering Dataset}
\paragraph*{\bf Answer sentence selection.}
QA tasks are classified by the level of structure in the context and the
type of the task. The answer can be {\em sentence-level} by selecting the relevant sentence from the context; {\em span-level} by choosing a span of the text from the context as the answer; or the answer can be generated using predicate values and sentence generation
models. This paper performs sentence selection from limited unstructured data based on the
\BibleQA\ dataset. Specifically, the input $(Q,A)$ contains a list of $M$ questions
$Q = (q_1, q_2,\ldots q_M)$, and a list of candidate answers
$A = (a_{1,1},a_{1,2},\ldots,a_{1,N_1};\ldots;a_{M,1},a_{M,2},\ldots,a_{M,N_M})$, where  $a_{j,k}$ is the $k$th candidate of the $j$th question. The output will be evaluated likelihood $p_{j,k}\in[0,1]$ for each $a_{j,k}$ to be the correct answer sentence for question $q_j$. Finally, the output answer $a_{j,k^*}$ corresponding to $q_t$ is the one with
the highest likelihood, i.e., $k^* = \arg\max_{k}\{p_{j,k}\mid 1\leq k\leq N_j\}$.

\paragraph*{\bf \SQuAD.} The Stanford Question
Answering Dataset (\SQuAD) is currently the largest span-based dataset, containing more
than $100,000$ QA pairs from more than 500 Wikipedia articles \cite{rajpurkar2016squad}. The dataset is span-based, meaning that given a context paragraph and a question, the dataset outputs the span of text that is the most likely to be the answer to the question.
Since we are interested in a sentence-level task, we converted \SQuAD\ to a new
sentence-level dataset. For each original context paragraph, we divide the paragraph into
sentences. Then, we label each sentence based on whether or not the
originally given span answer is within the sentence. If yes, then the sentence is labelled as 1,
and all the other sentences within the same paragraph are labelled as 0.

\paragraph*{\bf \BibleQA\footnote{{\tt https://github.com/helen-jiahe-zhao/BibleQA}}.}   Although the Bible was written over a period of 1000 years by 40 different people, it tells a unified story regarding the overarching theme of God's redemptive work in the past, present, and future. The Bible contains 39 books in the Old Testament and 27 books in the New Testament. The Old Testament contains a variety of literary genre, including historical narrative, wisdom books, poetry, and prophecy. The historical books describe the establishment of Israel as the chosen people of God and their separation from God due to sin, and the prophetic books prophesied about a coming Messianic King who will rescue humanity from the bondage of sin. 
The first part of the New Testament contains stories about Jesus' life and teachings, who was widely considered as the Messiah prophesied by the Old Testament prophets. The rest of New Testament consists of letters of instructions that are extensively read today and are foundational in the Christian teaching. The Bible is a collection of ancient religious literature that has undeniable influence over culture and society, affecting areas ranging
from languages, literature, to law and sciences \cite{daniell2003bible}.
It was found through a study that two-thirds of the American people believe that the Bible holds the answers to all or most of life's basic questions \cite{prothero2007religious}. From these, we can see the multi-faceted potential of the Bible being used in QA systems.

There is no existing dataset directly suitable for sentence-level QA task for the
Bible, and we set out to create our dataset using some existing available questions. We used a freely available set of 1001 trivia questions from the Bible\footnote{\texttt{https://biblequizzes.org.uk/}}
as the basis for the dataset.
The trivia question set consists of question, answer, and the corresponding verse from the Bible
which is relevant to the answer.

Using this resource, we derived a sentence-level dataset which we will name \BibleQA. We extracted verses surrounding the target verse as candidate answers. The actual verse is labelled with 1, and all the other verses are labelled with 0. There were also some questions from the original list that wasn't very suitable to answer using the Bible, especially questions that we cannot directly find answers from the Bible such as ``Which book of the Bible has the most chapters in it?''. Therefore, we filtered out those question and added more questions manually. We ended up with 886 questions in total in the \BibleQA\ dataset. An example is given below:

\smallskip

{\begin{description}
\item[Question:] What is the name of Jesus' mother?
\item[Verse 1:] [Matthew 1:17] \textsl{So all the generations from Abraham to David are fourteen
generations; from David to the exile to Babylon fourteen generations; and from the
carrying away to Babylon to the Christ, fourteen generations.}
\item[Verse 2:] [Matthew 1:18] \textsl{Now the birth of Jesus Christ was like this; for after his mother, Mary, was engaged to Joseph, before they came together, she was found pregnant
by the Holy Spirit.}
\item[Verse 3:] [Matthew 1:19] \textsl{Joseph, her husband, being a righteous man, and not willing to
make her a public example, intended to put her away secretly.}
\item[Answer:] $[0, 1, 0]$
\end{description}}

It is also noteworthy that there are many different English translations of the Bible, each
using a different translation philosophy and results in slightly different verses that consist of
the same idea. We decide to make use of digitalized versions of the Bible found on GitHub\footnote{\texttt{https://github.com/scrollmapper/bible\_databases}}.
We retrieve the verses from four public domain translations: King James Version (KJV), Young's
Literal Translation (YLT), American Standard Version (ASV) and World English Bible (WEB).
King James Version was published in 1611, and is still one of the most widely used Bible
translations and considered by many as the most authentic. Young's Literal Version, from
1862, follows a strictly literal translation philosophy and translated into English from Greek and
Hebrew almost word by word. American Standard Version from 1901 was very popular in the
20th century in its usage by biblical scholars. The World English Bible is an updated version
of the ASV and is the most modern translation out of all four being published in 2000. We
chose these four translations for their variety in the use of English and translation method. A
comparison of the translations is shown below. From the comparison we can see that each translation has its subtle differences, yet substantially different. For \BibleQA\ we used four translations for each question and answer pair,
which meant that the total size of the dataset ends up with 3544 question-answer pairs.

\smallskip

{
\noindent {\bf KJV:} \textsl{Now the birth of Jesus Christ was on this wise: When as his mother Mary
was espoused to Joseph, before they came together, she was found with child of
the Holy Ghost.} (Matthew 1:18)

\noindent {\bf YLT:} \textsl{And of Jesus Christ, the birth was thus: For his mother Mary having been
betrothed to Joseph, before their coming together she was found to have conceived
from the Holy Spirit.} (Matthew 1:18)

\noindent {\bf ASV:} \textsl{Now the birth of Jesus Christ was on this wise: When his mother Mary had
been betrothed to Joseph, before they came together she was found with child of
the Holy Spirit.} (Matthew 1:18)

\noindent {\bf WEB:} \textsl{Now the birth of Jesus Christ was like this; for after his mother, Mary, was
engaged to Joseph, before they came together, she was found pregnant by the Holy
Spirit.} (Matthew 1:18)

}

\section{Methodologies}
For our task, we employ three main neural network models for comparison purposes: one using recurrent neural network (RNN), one using convolutional neural networks (CNN), another using an adapted Bi-direction Attention Flow model (BiDAF) first suggested by Seo et al. \cite{seo2016bidirectional}.
The three models all follow the same general architecture, and the subsequent sections will describe the architecture in more detail:
\begin{enumerate}
\item {\em Embedding:} The input question and answers are first pre-processed and converted to
word vectors.
\item {\em Encoding:} The embedded sentences are then processed and encoded, to obtain one single
vector representation that captures the sentence. \item {\em Answer Selection:} Based on the encoded question and answer, select an answer as the
predicted output.
\end{enumerate}

\subsection{Word Embedding}
Word embedding captures word context using distributed word vectors. The underlying intuition is that words in similar environments tend to have similar meanings \cite{mikolov2013efficient}. Here we make use of both \GloVe\ vectors as well as \wordtvec.  \wordtvec\ is modelled as a shallow, two-layered neural network which uses stochastic gradient descent and back-propagation to iteratively make a word
embedding more similar to that of its neighbor words.
The model successfully reduces the complexity of the non-linear hidden layer and made it possible to learn high dimension word vectors on a significant amount of data.
\GloVe\ is an alternative unsupervised
learning algorithm to \wordtvec\ vectors, which is also used to obtain vector representation
for words \cite{pennington2014glove}. \GloVe\ has pre-trained vectors available online, that were trained on 6 billion tokens
from Wikipedia and various news outlets, making it very suitable for training on \SQuAD.

As the Bible consists of many words and names that are unique to its context, we also trained our own word vectors for the Bible. We used a combination of all four aforementioned English translations for the word vector training, which includes around 3 million words altogether. In the vector training, we used a context window size of 5 and the Continuous Bag-of-Words algorithm to train vectors of dimension 200. The resulting word vectors were able to capture the general semantics of the Bible-specific vocabularies.
Below are some of the most similar words for a few selected words in descending order of similarity, using the derived word vectors. From these lists, we see that the most similar words for `God' captures many qualities and
roles God is seen to have throughout the Bible. The most similar words for `David' are other
names who were closely related to him: Saul, his primary adversary; Absalom and Solomon,
his sons; Joab his army commander and Jonathan his best friend.

{\begin{description}
\item[God:] lord, saviour, holiness, mercy, lovingkindness, sworn, redeemer, salvation,
jehovah, endureth
\item[sin:] trespass, sins, guilt, guilty, transgression, forgiven, sinned, iniquity, forgive,
ignorance
\item[david:] saul, absalom, joab, abimelech, solomon, abner, jonathan, abraham, achish,
samuel
\end{description}}

The trained word vectors were concatenated with the \GloVe\ vectors in the transfer learning process so that the training on \BibleQA\ would be more meaningful.

\subsection{Models for the QA System}

\paragraph*{\bf The baseline model.} This model acts as the basis of comparison for all other results. The model uses a
random function to uniformly randomly generate an output in the range $[0,1]$ for each data point.
The baseline gives us a model that performs at a level
that does not involve any learning and simply assigning a random prediction for each question-
answer pair. The baseline is then compared with our models to evaluate the improvement made
by more sophisticated models.

\paragraph*{\bf The RNN model.} Recurrent networks are designed to model sequences, allowing the
users to work with sequences while preserving structural information. They are particularly
useful in NLP tasks due to the sequential nature of languages.
 A recurrent network consists of loops, where the output of a particular layer is passed back to the same layer as input. This allows information to persist and capture long-term dependencies such as those that appear in sequences.
One of the most popular implementations of RNN is the Long Short-Term Memory (LSTM), which was introduced to mitigate the
vanishing gradients problem \cite{hochreiter1997long}. As the sequence grows longer, the distance between the current
word and the dependent context grows longer. However, this means that the error gradients in
later steps in the sequence diminish quickly in the back-propagation and do not reach earlier
input signals, hence the gradients ``vanish''. This makes it very difficult to capture relevant
information. LSTM introduces a vector that acts as a memory cell, which preserves gradients
over time. The access to the memory cell is controlled by gating components that can be
thought of as logical gates.

Our RNN model makes use of LSTM layers to produce vector representations of the question
and answer phrases. The output is obtained as a probability between $[0,1]$ that indicates the
similarity between the question vector and the answer vector. This is based on the intuition that
sentences that have closer vectors should be more similar, and therefore
the answer should be more relevant to the question.

\begin{figure}
\centering
\resizebox{!}{2.7cm}{\includegraphics{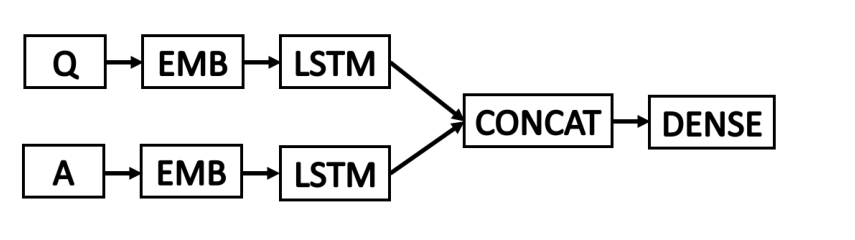}}
\caption{Recurrent Neural Network Model}\label{fig:RNN}
\end{figure}

The word embedding layer transforms each word into a word vector.
A question is then a sequence of word vectors $\vec{x}=(x_1,x_2,\ldots,x_t)$ and an answer is another sequence of word vectors $\vec{y}=(y_1,y_2,\ldots,y_{t'})$.
The encoding procedure applies two LSTMs, one for questions $Q$ and the other for answers $A$: For each example $(\vec{x},\vec{y})$, it sets for each $t$
\[
s_t=(C_t,h_t)=R_{\LSTM}(s_{t-1},x_{t-1}) \text{, and }
\]
\[
s'_t = (C'_t,h'_t)= R'_{\LSTM}(s'_{t-1},y_{t-1})
\]
where $s_t$ ($s'_t$) is the $t$th state, $C_t$ ($C'_t$) is the memory-cell states, and $h_t$ ($h'_t$) is the output state and $R_{\LSTM}$ ($R'_{\LSTM}$) is the LSTM networks for the question (answer). The output would be $\mathbf{m}=(Q_e,A_e)$ where $Q_e = (h_1,h_2,\ldots,h_t)$, $A_e = (h'_1,h'_2,\ldots,h'_t)$. Finally, we concatenate the question and answer vectors,  pass them through a final layer which uses the sigmoid activation function $\sigma=\frac{1}{1+e^{-x}}$, and obtain predicted likelihood for the answer being the correct one for the question.
Fig.~\ref{fig:RNN} shows an overview of the layers.

\paragraph*{\bf The CNN model.}  CNNs are special feed-forward neural networks
with fully connected layers and consisting of convolutional and pooling layers.
These specialized layers are useful for finding strong local information present within the input regardless of the position of the signals, e.g., in a QA task, a sentence may contain a key phrase that strongly indicates it as the answer to a question.
In NLP, a CNN network first takes a sequence of words and applies a
filter over each $n$-gram of the sequence obtained by a sliding window of $k$ words. The filter
transforms the $n$-gram into a $d$-dimensional vector that captures important properties of the
words. Finally, the pooling layer combines all the $d$-dimensional vectors into one single $d$-dimensional vector by taking a max or average operation over each dimension of the vector.
This final vector is then used for further processing in the neural network since it now contains
some of the most important local information in the entire sequence.

Our CNN model uses the convolutional and pooling layers to represent the question and answer
phrases.
We also use a dropout layer to regulate the weights and avoid overfitting.
For each question and answer sequence, the convolution layer
applies a kernel across the sequence, transforms it using a filter, passes through a max-pooling
layer to obtain the element-wise maximum, and finally passes through an output layer which
returns a prediction.

\begin{figure}
\centering
\resizebox{!}{3.5cm}{\includegraphics{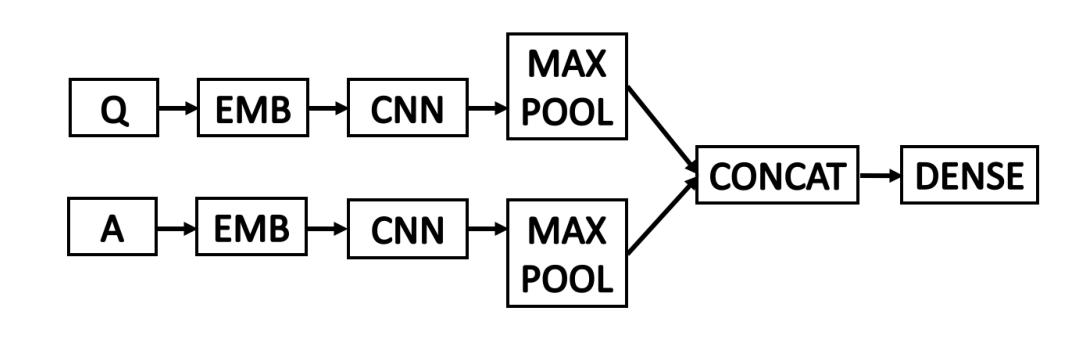}}
\caption{Convolutional Neural Network Model}\label{fig:CNN}
\end{figure}

The convolution and max-pooling layers for each question $\vec{x}$ and answer $\vec{y}$ is:
\[
Q^c_i=f\left(W^\mathrm{T} \vec{x}_{i:i+k-1} + \vec{b}\right)
\text{ and }
A^c_i=f\left(W^\mathrm{T} \vec{y}_{i:i+k-1} + \vec{b}\right)
\]
\[
Q^v_k = \max_{1<i<m} Q^c_i[k] \text{ and }A^v_i= \max_{1<i<m} A^c_i[k]
\]
where $k$ is the window size, $f$ is the \textsf{relu} activation function $f(x) = \max\{0,x\}$, $W$ is the filter vector that performs the linear transformation and  $\vec{b}$ is the bias parameter of the network.
Finally, the output layers are the same as for the RNN model. Fig.~\ref{fig:CNN} shows an overview of the layers.

\paragraph*{\bf BiDAF Model.}
The BiDAF model  was proposed for the \SQuAD\ dataset for span-level QA
tasks \cite{seo2016bidirectional}. The original model was used
 to find the start
and end indices of the answer to a question within a context paragraph:
The question and context paragraphs are first converted to vectors using both word and character embeddings then combined to form a phrase embedding using LSTM.
The question and context paragraphs are then combined to produce a set of query-aware vectors for every word in the context. Finally, another LSTM is used to scan the context paragraph, and the output layer produces the start and end indices for the question.

\begin{figure}
\centering
\resizebox{!}{2.5cm}{\includegraphics{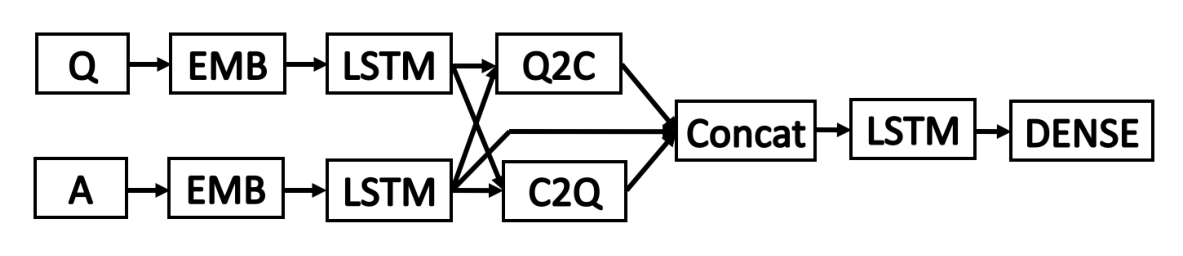}}
\caption{Modified Bidirectional Attention Flow Model}\label{fig:BiDAF}
\end{figure}

For our sentence-level task, we modify the original BiDAF model slightly. We use the candidate sentences as the context in the original model, eliminate the use of character-embeddings, and output only the probability of the answer sentence being the correct answer to the question. Fig.~\ref{fig:BiDAF} shows an overview of the layers.

The word and phrasal embedding layers are similar to the RNN model in using LSTM, whose result contains the matrix representations of a question $\mathbf{q}$ and a candidate answer $\mathbf{a}$. Following
that, Q2C and C2Q are two layers that compute the ``attention'' for the question and answer
which is essentially the interaction between question and answers. More formally, we use $U_{:i}$ to indicate the $i$th column vector of any matrix $U$. The bidirectional attention
is determined by a {\em similarity matrix} $S_{j,k}$ between the $j$th answer word and the $k$th question word,
defined as $S_{j,k} = \alpha(\mathbf{a}_{:j},\mathbf{q}_{:k})$
where $\alpha(\vec{a},\vec{q})$
is a trainable scalar function that represents the similarity between vectors $\vec{a},\vec{q}$.
The Q2C, or query to context layer, signifies which answer words have the closest similarity
to one of the question words.
The C2Q, or context to query layer, determines which question words are the most relevant
to each answer word. The exact definition of $S_{j,k}$ and Q2C, C2Q can be found in~\cite{seo2016bidirectional}.

The Q2C and C2Q are then concatenated with the answer embedding to output query-aware vector representations of the context words; it takes the form of a matrix $G$, whose $j$th column is
\[
G_j = \beta(A_{:j}, \tilde{Q_{:j}}, \tilde{A_{:j}})
\]
where $\beta$ is a trainable function that combines the three representations.
The result is then passed on to one more LSTM layer to output representation matrix $\mathbf{m}$, and finally to the output
layer which predicts the probability of $\mathbf{a}$ being the answer of $\mathbf{q}$.

\section{Experiments}
\subsection{Experiment Setup}

We performed experiments for each of the baseline, RNN, CNN and BiDAF model. For any dataset, we use 70\% of
the samples for training and 30\% for testing. Out of the 70\% training
data points, 10\% will be used as the validation set for monitoring overfitting. At each epoch,
the model trains on the training sets and calculates the loss and accuracy concerning both the
training and validation set. If the validation loss is much higher than the training loss and the
validation accuracy is much lower than training accuracy, then we can conclude that the model
is overfitting to the training set and will need to modify our model accordingly.

We used the \GloVe\ word vectors (dimension $d = 100$) for training the \SQuAD\ models, and  both \GloVe\ and
\wordtvec\ vectors for the \BibleQA\ models where the \wordtvec\ vectors ($d = 200$)
are trained on 4 versions of the Bible.
The loss function that our models will learn to minimize is the {\em binary cross entropy}, defined as:
\[
L(\Theta) = -\frac{1}{N} \sum_{i=1}^N (y_i \cdot \log(p_i) + (1 - y_i ) \cdot \log(1 - p_i ))
\]
where $\Theta$ is the set of parameters, $N$ represents the number of training instances,
$p_i$ is the probability of class 1, $1-p_i$ is the probability of class 0, and $y_i\in\{0,1\}$ is the true label of the
$i$th observation. Here, the value of $p_i$ is found by the probability of the activation layer using a sigmoid activation: For weight vector $\Theta$ and input vector $\vec{x}$, $p_i = \frac{1}{1+e^{-\Theta \cdot \vec{x}}}$.

We used the {\em adaptive gradient} (AdaGrad) optimizer to train the neural networks, which
is a modified stochastic gradient with per-parameter learning rate \cite{duchi2011adaptive}. The learning rate of
a model determines the rate of update between each iteration of backpropagation. AdaGrad allows the learning rate to adapt based on the parameters. It performs larger updates for infrequent parameters and smaller updates for frequent parameters, and often improves convergence in tasks where the data is sparse -- such as NLP and image recognition.
Let $g_\tau = \nabla L(\Theta)$ be the gradient at iteration $\tau$. The per-parameter update for AdaGrad
uses the following formula:
\[
\Theta_{\tau+1} = \Theta_\tau - \frac{\eta}{\sqrt{G_{\tau,\tau}}}g_\tau
\]
where $\eta$ is the learning rate, and $G_{\tau,\tau} = \sum_{j=1}^\tau g^2_{j,\tau}$
produces a scaling factor for the
parameter $\Theta_\tau$. This leads to a different learning rate update for each parameter
based on the scaling factor and the learning process is adaptive.

For our experiments, we will use two metrics to evaluate their performance: (1) {\em F1 score}, a widely-used accuracy indicator, is defined as $F_1=2PR/(P+R)$ where $P$ and $R$ are precision and recall, resp. (2) {\em Mean reciprocal rank }(MRR) which commonly measures accuracy of ranked outputs, and is applicable here as we are essentially ranking the candidate sentences for each question. MRR is defined as $\frac{1}{n}\sum_{i=1}^{n} \mathsf{rank}_i^{-1}$ where $n$ is the number of questions and $\mathsf{rank}_i$ is the rank of the correct answer of the $i$th question.

\subsection{Experiment 1: Transfer Learning Parameter Tuning}
\noindent {\bf Goal and method.} The first experiment investigates the effect of transfer learning on the model accuracy. We pre-train the model on the \SQuAD\ dataset and compare that with training only on \BibleQA.
For each model, 1) we first run the model on \BibleQA\ to obtain a set results. 2) Then, we train the same model on \SQuAD\ to obtain the trained weights. 3) Finally, we run the model on \BibleQA\ once
again using the trained weights from \SQuAD\ and perform weight fine-tuning once again. We
then compare the results of each model to see whether there are improvements from
before using the transferred weights. We tune parameters such as learning rate and epoch
to find the best performing model. We find that a learning rate of $\eta = 0.001$
worked the best for the RNN and BiDAF model, and $\eta = 0.0001$ for the CNN model. We use
an early stopping mechanism for determining the optimal number of epochs trained, which
monitors the validation loss at each epoch and stops the training once the model stops improving.
We set the patience to 10, which means that the model will wait for 10 epochs before
terminating the training. Optimal results are achieved with 20 to 30 epochs.

\paragraph*{\bf Results and analysis.} Table~\ref{tab:transfer} contains the results we obtained from before and after the weight transfer. We can see that using the transferred weight improves the F1 results by
0.08 to 0.09 (which is around $20\%$-$30\%$ improvement). This shows that just as we hypothesized, pre-training had a positive effect on the training accuracy. However, while the F1 score increases with transferring weights, the MRR
of the model decreased by 0.05 and 0.06 for the CNN and BiDAF model, with only the RNN model increasing by 0.05. This was surprising as it is often assumed that there is a correlation
between different evaluation measures -- that a higher F1 would also result in a higher MRR.
By considering what the MRR is measuring, it seems that these models had a higher average
ranking for the correct output. However since these models also had lower F1, they are less
likely to choose the correct answer as the top ranking answer. So while the models improved the F1 score, it is choosing the correct output more often, but it also ranks the
correct answer lower in the cases where the model incorrectly predicts the results. This is an
interesting phenomenon and will be worth looking into in the future. Out of the three models, the RNN model performs the best overall with the highest F1
score both before and after weight transfer, and the highest MRR after weight transfer.

\begin{table}\caption{Model Comparison Before and After Weight Transfer}\label{tab:transfer}
\centering
\begin{tabular}{|c|c|c|c|}\hline
Model&Transferred Weights& F1 & MRR\\\hline \hline
Baseline &No &0.35& 0.59 \\ \hline
\multirow{2}*{RNN} & No & 0.45 & 0.56 \\ \cline{2-4}
    & Yes& {\bf 0.54} & {\bf 0.61} \\ \hline
\multirow{2}*{CNN} & No & 0.39 & 0.58 \\ \cline{2-4}
    & Yes& 0.48 & 0.53 \\ \hline
\multirow{2}*{BiDAF}&No & 0.40 & 0.59\\ \cline{2-4}
     &Yes& 0.48 & 0.53\\ \hline
\end{tabular}
\end{table}

\subsection{Experiment 2: Answer Context Length}
\noindent {\bf Goal and method.} The second experiment aims to find the variation of prediction results by changing the length of the answer context, or in other words, the number of candidate sentences for each question. During
the dataset construction phase, after we identify the verse that corresponds to the correct answer to a
question, we include a different number of context verses surrounding the correct answer.
We created three types of datasets this way: \BibleQA-3, \BibleQA-10, and \BibleQA-chapter.
For \BibleQA-3 and \BibleQA-10, we included 3 and 10 verses surrounding the true verse respectively as candidates. For the chapter version, we included all verses from the same chapter that usually ranges from 10 to 60 verses. Each RNN, CNN, and BiDAF model was tuned on \BibleQA-3 for the maximal accuracy result, and subsequently, the same model was used for the prediction for \BibleQA-10 and
\BibleQA-chapter. For all datasets, we included all four Bible translations: KJV, ASV, YLT and WEB.

\paragraph*{\bf Results and analysis.} We used the tuned models from the last experiment for training each dataset, which compares the effect of changing the length of the context has on the model accuracy. Table~\ref{tab:context} describes the results among the three datasets with different answer
context lengths. Across all three datasets, the CNN model performs the best using the F1 measure, while BiDAF generally has the best MRR score. This echoes the interesting
phenomenon as mentioned above as to why certain models would have a higher F1 score but
lower MRR than others. Once again, more investigation is needed.

For the shortest context with three verses, all the models significantly improve on the
baseline by 0.13 to 0.19 F1, with the best results from the RNN model on both F1 and MRR. As
the context length increases to 10 verses, the model accuracy significantly decreases, improving
only 0.03 to 0.05 of the F1 score from the baseline model. The RNN drops its performance
compared to others, and CNN rises as the model with the highest F1, but BiDAF becomes the
model with the highest MRR. Finally, in the longest context length of using the entire chapter,
the models perform at around the same level as the baseline model, if not worse.

This shows that the models are not yet able to be used for longer contexts. A larger dataset and longer training time could be used to train a more robust model that can
deal with longer context lengths in the future.

\begin{table}\caption{Comparison between \BibleQA-3, \BibleQA-10 and \BibleQA-Chapter for different
models}\label{tab:context}
\centering
\begin{tabular}{|c|c|c|c|}\hline
Dataset&Mode& F1 & MRR\\\hline \hline
\multirow{4}*{\BibleQA-3}& Baseline &0.35& 0.58 \\ \cline{2-4}
& RNN &{\bf 0.54}& {\bf 0.61} \\ \cline{2-4}
& CNN &0.48& 0.53 \\ \cline{2-4}
& BiDAF &0.48& 0.53 \\ \cline{1-4}
\multirow{4}*{\BibleQA-10}& Baseline &0.11& 0.30 \\ \cline{2-4}
& RNN &0.14& 0.30 \\ \cline{2-4}
& CNN &{\bf 0.16}& 0.28 \\ \cline{2-4}
& BiDAF &0.14& {\bf 0.32} \\ \cline{1-4}
\multirow{4}*{Chapter}& Baseline &{\bf 0.05}& {\bf 0.15} \\ \cline{2-4}
& RNN &0.02& 0.14 \\ \cline{2-4}
& CNN &{\bf 0.05}& {\bf 0.15} \\ \cline{2-4}
& BiDAF &0.04& 0.14 \\ \cline{1-4}
\end{tabular}
\end{table}

\subsection{Experiment 3: Translation Version}
\noindent {\bf Goal and method.} The third experiment focuses on the differences among various English translations of the Bible. As mentioned above, we used 4 English translations in training
the word vectors creating the dataset. Each of these translations varies in their translation philosophy, as well as the modernity of their language. YLT is the most literal English translation of
the original languages. The other three translations are roughly the same in their translation
philosophy in that they are not as literal as YLT, but also strive to truthfully capture the original meaning. Ordering them by modernity, KJV was the oldest translation
being published 400 years ago, while WEB is the most recent at
2000. The literality of the translations could affect the sentence representations, as they could
choose to use certain words for translation that are a more direct translation of the intended
meaning. The modernity of the language could also affect the word vector, since
older translations are likely to contain obsolete words that are no longer used, and therefore
the word vectors may not necessarily capture an accurate representation. We want to compare whether the model prediction changes based on the level of literal translation, or by the language modernity.
For this experiment, we create four further datasets, each only using one particular translation. We use the format of \BibleQA-10, and select 10 candidate verses for each question. We use the same three models, and compare the result for each translation individually.

\noindent {\bf Results and analysis.} Table~\ref{tab:translation} compares the results for each translation within each model. Looking at both F1 and MRR scores,  WEB achieves the best performance, having the top MRR result for RNN and the top F1 score for CNN/BiDAF.
This suggests that using a translation with more modern language can be
beneficial for the QA process, and it could be the case that the word vectors
used were able to capture more accurate meanings.

The KJV translation follows the WEB translation and has the highest performance for
the RNN model using the F1 measure, as well as for the CNN model under MRR. The high
performance was surprising, as we expected that for a translation such as KJV, some of the
archaic language used in the translation could have been a deterrent for learning useful word
vectors. It turns out that despite the choice of English words, the KJV still may perform relatively
well in NLP tasks.

The YLT has the highest MRR result for the BiDAF model. However, the variance is not
large enough for the result to be significant, and YLT does not perform particularly well in any
other models. From this, we conclude that the translation philosophy and the level of literalness
do not necessarily play a dominant role in training QA system.

\begin{table}\caption{ Translation comparison for each model. The bolded figures represent the highest
score for each model}\label{tab:translation}
\centering
\begin{tabular}{|c|c|c|c|}\hline
Dataset&Mode& F1 & MRR\\\hline \hline
\multirow{5}*{RNN}& KJV &{\bf 0.17}& 0.30 \\ \cline{2-4}
& ASV &0.14& 0.29 \\ \cline{2-4}
& YLT &0.09& 0.31 \\ \cline{2-4}
& WEB &0.12& {\bf 0.32} \\ \cline{2-4}
& Combined &0.14& {\bf 0.32} \\ \cline{1-4}
\multirow{5}*{CNN}& KJV &0.13& {\bf 0.35} \\ \cline{2-4}
& ASV &0.14& 0.28 \\ \cline{2-4}
& YLT &0.14& 0.29 \\ \cline{2-4}
& WEB &{\bf 0.16}& 0.26 \\ \cline{2-4}
& Combined &0.16& 0.28 \\ \cline{1-4}
\multirow{5}*{BiDAF}& KJV &0.11& 0.28 \\ \cline{2-4}
& ASV &0.10& 0.29 \\ \cline{2-4}
& YLT &0.11& 0.30 \\ \cline{2-4}
& WEB &{\bf 0.16}& 0.28 \\ \cline{2-4}
& Combined &0.14& {\bf 0.32} \\ \cline{1-4}
\end{tabular}
\end{table}

\section{Conclusions and Future Work}
In this paper, we leverage transfer learning techniques to study domain adaptation in QA tasks using the \BibleQA\ dataset. Transferring the weights from the much larger $\SQuAD$ dataset has a noticeable improvement in the model accuracy. This showed the potential of using transferred weights for this particular task. We also find that RNN was the best performing model, while BiDAF did not perform as well as expected
despite being the most complicated model. This suggests that simpler architectures can still
sometimes achieve relatively good results.

When increasing the number of candidate sentences to choose from as answers to questions, unsurprisingly, the model performances deteriorates. Comparing different Bible translations that vary in degrees of literalness in translation as well as the modernity of language, we find that the World English Bible gives the best results, followed by the King James Version. The modernity of the language may be attributed to the good performance of WEB. At the same time, although KJV was written centuries ago and uses different words than we do now, it is still able to produce useful results. Furthermore, Young's Literal Translation being the most literal translation does not perform particularly well. We conclude that the translation philosophy, and how literal a translation is, does not necessarily improve the results.

Our system has certain limitations, and here we will suggest some potential improvements
and direction for future research. The implementation of the BiDAF model was entirely dependent on the DeepQA library, which has very recently been deprecated. The researchers behind
DeepQA has ported the library to PyTorch\footnote{ http://pytorch.org/}
which they have found to be better for NLP research. In the future, it could be worthwhile to consider implementation using PyTorch instead of Keras, and to use reliable and stable software frameworks. The sentence encoding methods
used in our models are still relatively simple, in particular, the RNN and the CNN models. The
main issue with our current method of encoding is that it mostly only takes into consideration
the semantics of the sentences, and not as much the syntax. While it is a simple method that
has shown to have worked relatively well, to achieve better accuracy we could consider incorporating an encoding scheme which considers both the syntax and the semantics, such as the treeLSTM \cite{tai2015improved} or an ensemble of different encoding schemes.
The domain adaptation methods used in our systems was also a simple approach which
only involves pre-training the weights and transferring the weights. More exploration into
improving the transfer learning method could be beneficial. For example, transferring only
the weights of certain layers or tuning them at different learning rates. As transfer learning
becomes more widely used in NLP research, we expect that more effective methods would
emerge that can improve the system. The accuracy of the model is highly dependent on the
quality of the dataset. The \BibleQA\ dataset we created was had only 886 distinct
question. Extending the size of the dataset is also a worthwhile task in the future, that can be done by manually adding more questions, combining with other sources of Bible questions, or potentially leveraging
techniques that could automatically generate questions based on a text.

\bibliographystyle{plain}

\end{document}